\documentclass[twocolumn]{revtex4-1}

\usepackage{graphicx}
\usepackage{txfonts}
\usepackage{color}

\begin{document}

\title{Can the excess in the FeXXVI Ly$\gamma$ line from the Galactic Center
provide evidence for 17 keV sterile neutrinos?}

\author{D. A. Prokhorov}
\email{phdmitry@gmail.com}

\affiliation{Korea Astronomy and Space Science Institute,
Hwaam-dong, Yuseong-gu, Daejeon, 305-348, Republic of Korea}

\author{Joseph Silk}
\affiliation{Astrophysics, Department of Physics, University of
Oxford, Keble Road Ox1 3RH, Oxford, United Kingdom}

\affiliation{Institut d'Astrophysique de Paris, 98bis Blvd Arago,
Paris 75014, France}

\begin{abstract}

{The standard model of particle physics assumes that neutrinos are
massless, although adding non-zeros is required by the
experimentally established phenomenon of neutrino oscillations
requires neutrinos to have non-zero mass. Sterile neutrinos (or
right-handed neutrinos) are a good warm dark matter candidate. We
find that the excess of the intensity in the 8.7 keV line (at the
energy of the FeXXVI Ly$\gamma$ line) in the  spectrum of the
Galactic center observed by the Suzaku X-ray mission cannot be
explained by standard ionization and recombination processes. We
suggest that the origin of this excess is via decays of sterile
neutrinos with a mass of 17.4 keV and estimate the value of the
mixing angle. The estimated value of the mixing angle
$\sin^2(2\theta)=(4.1 \pm 2.2)\times10^{-12}$ lies in the allowed
region of the mixing angle of dark matter sterile neutrino with a
mass of 17-18 keV. }
\end{abstract}

\maketitle

\section{Introduction}

Several astrophysical observations, such as the indications of
central cores in low-mass galaxies, the low number of satellites
observed around  the Milky Way, and the near constant cores of the
least luminous satellites, have revived interest in sterile
neutrinos as a  warm dark matter candidate. Electroweak singlet
right-handed (sterile) neutrinos  with masses in the few keV range
naturally arise in many extensions of the Standard Model and could
be produced in the early universe through the Dodelson-Widrow
mechanism  involving (non-resonant) oscillations with active
neutrino species \cite{Dodelson 1994}.  The experimentally
established phenomenon of neutrino oscillations requires neutrinos
to have non-zero mass, and sterile neutrinos (or right-handed
neutrinos) are a natural warm dark matter candidate.

Direct constraints on masses and mixing angles are  obtained both
from the Lyman alpha forest power spectrum and  X-ray observations
of the radiative decay channel, the latter providing a photon with
energy E=$m_{\mathrm{s}} c^2/2$, where $m_{\mathrm{s}}$ is the
sterile neutrino mass. Most recently, X-ray observations of the
local dwarf Wilman 1 have shown marginal evidence for a 5 keV
sterile neutrino \cite{Loewenstein 2009}. The inferred mixing angle
lies in a narrow range for which neutrino oscillations can produce
all of the dark matter and for which sterile neutrino emission from
cooling neutron stars can explain pulsar kicks.

In fact if  sterile neutrinos provide a significant fraction
(although not necessarily all, see \cite{Palazzo 2007},) of the dark
matter, the Galactic Center provides an even more attractive
environment to search for radiative decay signals.  If the sterile
neutrino mass is indeed around 17 keV, we expect a X-ray line near
$8.5$ keV. The diffuse X-ray emission from the Galactic Center using
the X-ray imaging spectrometer on Suzaku was analyzed by
\cite{Koyama 2007}, who detect the 8.7 keV line corresponding the
FeXXVI Ly$\gamma$. We have reanalysed the data on hydrogen-like iron
line strengths and find that there is an unexpected excess in the
8.7 keV line from the Galactic center in the Suzaku data. We
demonstrate that this excess cannot be explained by means of
standard ionization and recombination processes. We propose an
explanation of the excess in the 8.7 keV line in terms of  17.4 keV
neutrino decays.

\section{The excess in the 8.7 keV line from the Galactic center and its origin}

X-ray emission from the Galactic center has been observed for almost
30 years. A component of the diffuse  emission is thermal   and is produced by a high
 temperature plasma in the inner 20 parsecs of the Galactic Center
($\sim$ 8 keV). The most pronounced features in the emission lines
are Fe I K$\alpha$ at 6.4 keV, and the K-shell lines 6.7 and 6.9 keV
from the helium-like (FeXXV K$\alpha$) and hydrogen-like (FeXXVI
Ly$\alpha$) ions of iron, respectively. An analysis of the ratio of
the 6.7 keV to 6.9 keV lines is an interesting test of plasma
components (see, e.g. \cite{Dogiel 2009}, \cite{Prokhorov 2009}).
For the first time, the FeXXVI Ly$\gamma$ at 8.7 keV was detected by
the Suzaku X-ray mission \cite{Koyama 2007}. The observed line
intensities by Suzaku are listed in Table 2 of \cite{Koyama 2007}.
For convenience, we list below the most important lines (for our
analysis) taken from the paper by \cite{Koyama 2007}. The measured
intensities of the lines of the hydrogen-like iron ions are:
$I_{\mathrm{Ly}\alpha}=1.66^{+0.09}_{-0.11}\times 10^{-4}$
ph/(cm$^2$ s), $I_{\mathrm{Ly}\beta}=2.29^{+1.35}_{-1.31}\times
10^{-5}$ ph/(cm$^2$ s), and
$I_{\mathrm{Ly}\gamma}=1.77^{+0.62}_{-0.56}\times 10^{-5}$
ph/(cm$^2$ s). The errors are at 90\% confidence level \cite{Koyama
2007}.

The measured ratio of the FeXXVI Ly$\beta$ to FeXXVI Ly$\alpha$ iron
lines equals $\approx 0.138\pm 0.059$  and is in  agreement with the
theoretical value of $\approx 0.14$ in the gas temperature range
between 5 and 15 keV (see line list \cite{linelist}; for a review,
see \cite{Mewe 1981}).

We note that the measured intensity of the FeXXVI
Ly$\gamma=1.77^{+0.62}_{-0.56}\times 10^{-5}$ ph/(cm$^2$ s) iron
line has an significant excess above the value derived from the
theoretical model \cite{linelist} and the measured  intensity
$I_{\mathrm{Ly}\alpha}$ of the FeXXVI Ly$\alpha$ iron line. The
ratio of the the FeXXVI Ly$\gamma$ to FeXXVI Ly$\alpha$ iron lines
equaled $0.038$ in the gas temperature range between 5 and 15 keV is
from the theoretical model (see, \cite{linelist}). Therefore the
expected value of the intensity in the FeXXVI Ly$\gamma$ line is
$0.038\times I_{\mathrm{Ly}\alpha}=6.3^{+0.4}_{-0.4}\times10^{-6}$
ph/(cm$^2$ s) and is much smaller than the measured intensity by
Suzaku. Therefore, the excess of the intensity in the 8.7 keV line
equals $\approx (1.1\pm0.6)\times10^{-5}$ ph/(cm$^2$ s).

To demonstrate that this excess cannot be explained by ionization
and recombination processes, we calculate the ratio of the fully
stripped (FeXXVII) ionic fraction to the hydrogen-like iron ionic
fraction using the ratio of the intensities of the FeXXVI Ly$\gamma$
to Ly$\alpha$ lines as a function of temperature $T$.

The ratio $r_{\gamma\alpha}$ of the FeXXVI Ly$\gamma$ to FeXXVI
Ly$\alpha$ iron line intensities is given by
\begin{equation}
r_{\gamma\alpha}=\frac{E_{\gamma}(T) N_{\mathrm{FeXXVI}} +
R_{\gamma}(T) N_{\mathrm{FeXXVII}}}{E_{\alpha}(T)
N_{\mathrm{FeXXVI}} + R_{\alpha}(T) N_{\mathrm{FeXXVII}}} \label{r}
\end{equation}
where $E_{\gamma}$ and $E_{\alpha}$ are the impact excitation rate
coefficients, and $R_{\gamma}$ and $R_{\alpha}$ are the rate
coefficients for the contribution from recombination to the
Ly$\gamma$ and Ly$\alpha$ spectral lines, respectively. Excitation
and recombination rate coefficients are taken from \cite{Mewe 1981}.

From Eq. (\ref{r}) the ratio of the fully stripped (FeXXVII) ionic
fraction to the hydrogen-like iron ionic fraction is
\begin{equation}
\frac{N_{\mathrm{FeXXVII}}}{N_{\mathrm{FeXXVI}}}=\frac{E_{\gamma}(T)-r_{\gamma\alpha}
E_{\alpha}(T)}{r_{\gamma\alpha} R_{\alpha}(T)-R_{\gamma}(T)}
\end{equation}
Note that we do not use the assumption of  collisional ionization
equilibrium.

For the best fit value of the intensity ratio of the Ly$\gamma$ to
Ly$\alpha$ iron lines
($1.77\times10^{-5}/(1.66\times10^{-4})\approx0.107$) found by
\cite{Koyama 2007}, we find that the ratio of the fully stripped
(FeXXVII) ion fraction to the hydrogen-like iron ion fraction lies
in the range $(-40, -20)$ when the electron temperature is in the
range (5 keV, 15 keV). Since the ratio of the fully stripped
(FeXXVII) to hydrogen-like (FeXXVI) ion fractions should not be
negative, we conclude that the excess cannot be explained by
ionization and recombination processes. A more physically acceptable
explanation of the excess is that of  sterile neutrino decays, under
the assumption that sterile neutrinos constitute a significant
fraction of dark matter. In the next section, we calculate the
mixing angle using our inferred value of the intensity excess in the
8.7 keV line.

\section{Radiative decays of sterile neutrinos}

The sterile neutrino possesses a radiative decay channel decaying to
an active neutrino and a photon with energy $E=m_{\mathrm{s}}
c^2/2$. Since the excess in the 8.7 keV line corresponds to a
sterile neutrino mass of $17.4$ keV, in this section we estimate the
mixing angle $\sin^2(2\theta)$ and decay rate  for such a sterile
neutrino.

Using the excess in the 8.7 keV line intensity equal to
$I_{\mathrm{excess}}\approx (1.1\pm0.6)\times10^{-5}
\mathrm{ph/(cm^2 s)}$ and the definition of the energy flux
$F_{\mathrm{s}}=I_{\mathrm{excess}}\times E$, we find that the value
of the energy flux excess in the 8.7 keV line equals
$(9.6\pm5.2)\times10^{-5} \mathrm{keV/(cm^2 s)}$.

{Decays of sterile neutrinos in the dark matter halo of the
Milky Way are a promising way to explain the excess in the 8.7 keV
line. The amount dark matter within the field of view of the Suzaku
observation (see Fig. 1 of \cite{Koyama 2007}) is only a minute
fraction of the total mass of the dark matter halo of the Milky Way.
We consider the model of the Milky Way dark halo presented in
\cite{Klypin 2002} and used to derive constraints on the sterile
neutrino parameters by \cite{Boyarsky 2006}. The Milky Way halo
density is described by the Navarro-Frenk-White (NFW) profile
\begin{equation}
\rho_{\mathrm{NFW}}(r)=\frac{M_{\mathrm{vir}}}{4\pi\alpha}\frac{1}{r
(r_{\mathrm{s}}+r)^2},
\end{equation}
where the dark matter halo parameters of preferred models obtained in
\cite{Klypin 2002} correspond to $M_{\mathrm{vir}}=10^{12} \rm
M_{\odot}$, $r_{\mathrm{s}}=21.5$ kpc and numerical constant
$\alpha\simeq1.64$ (see \cite{Boyarsky 2006}). A mass within the
field of view of $M_{\mathrm{fov}}\simeq2.5\times10^6
\mathrm{M}_{\odot}$ is derived from this model.}

The decay flux into a solid angle $\Omega$ (within the field of
view) is given by (see, e. g. \cite{Boyarsky 2008})
\begin{equation}
F_{\mathrm{s}}=\int_{\Omega} \frac{\rho_{\mathrm{NFW}}(\vec{r})}
{4\pi (\vec{r}_{\mathrm{0}}-\vec{r})^2}\frac{\Gamma c^2}{2}
d^3\vec{r},
\end{equation}
{where $|\vec{r}_\mathrm{0}|=8.5$ kpc is the distance to the
Galactic center.} {The decay rate $\Gamma$ is given by (e.g.
\cite{Boyarsky 2008} and references therein)}
\begin{equation}
\Gamma=1.38\times10^{-22}
\sin^2(2\theta)\left(\frac{m_{\mathrm{s}}}{1
\mathrm{keV}}\right)^{5} \mathrm{s}^{-1}.
\end{equation}
To estimate the value of the mixing angle, we use the inferred value
of $m_{\mathrm{s}}=17.4 \mathrm{keV}$ and the value of the energy
flux excess. Then, from Eqs. (4), (5) the value of the mixing angle
is given by
\begin{equation}
\sin^2(2\theta)=(4.1\pm 2.2)\times10^{-12}.
\end{equation}

{Using Eq. (5) we derive the following constraint on the decay
rate:}
\begin{equation}
\Gamma=(9.0\pm4.8)\times 10^{-28} \mathrm{s}^{-1}.
\end{equation}

{Note that the derived values of the mixing angle of
$\sin^2(2\theta)=(4.1\pm2.2)\times10^{-12}$ and decay rate of
$\Gamma=(9.0\pm4.8)\times 10^{-28} \mathrm{s}^{-1}$ lie in the
allowed region for a dark matter sterile neutrino with a mass of
17-18 keV (see \cite{den Herder 2009}). Such a neutrino is capable
of accounting for a substantial fraction, if not all, of the dark
matter.}

{An additional contribution of comparable strength comes in a
point-like source if a cusp of dark matter develops around the
centar black hole. The dark matter cusp mass  is constrained
observationally by stellar orbit measurements and is taken to be
$3\times10^6\rm M_\odot $ within the central 0.001 pc. This is
comparable to the mass of the central black hole. Such a cusp is
naturally produced via adiabatic contraction of dark matter around
the black hole within its sphere of influence, about 3 pc for the
Milky Way black hole \cite{Gondolo 1999}, \cite{Zhao 2005}.
Typically about as much mass is captured in the spike as is in the
central black hole and with a density profile of $\rho\propto
r^{-9/4}$ or even steeper. Dynamical histories of merging complicate
this picture. In particular, binary black hole mergers, had they
occurred, would destroy such a spike (see \cite{Ullio 2001},
\cite{Merritt 2002}, \cite{Bertone 2005}).

However an alternative and more conservative scenario is cusp regeneration.
This cannot occur directly in the dark matter because its relaxation
time is extremely long. However, this is not the case for the
central stars, and once a cusp forms in the stars, scattering of
dark matter particles off of stars can redistribute the dark matter
in phase space on a time scale of order the relatively short
star-star relaxation time \cite{Merritt 2004}, resulting in a
$\rho\propto r^{-3/2}$ density cusp in the central parsec of the dark matter \cite{Merritt
2007}.

Sterile neutrinos would form such a dark matter cusp, although the spike
density is limited by phase space considerations for non-degenerate
neutrinos (see, \cite{Tremaine 1979}). Using the values of
the velocity dispersion for the Milky Way of $100$ km/s
\cite{Merritt 2001} and a sterile neutrino mass of 17.4 keV, we find
a limit on the density $\rho_{\rm cr}<1.4\times10^9 \rm
M_{\odot}/(pc^3)$. This would guarantee a point-like source if there is indeed a cusp.

}

\section{Conclusions}

We have found that there is an  excess in the intensity in the 8.7 keV line
observed from the Galactic center over the theoretical value expected for  the
intensity in the FeXXVI Ly$\gamma$ line. This intensity excess
equals $(1.1\pm0.6)\times10^{-5}$ ph/(cm$^2$ s), where the errors
are at the 90\% confidence level.
We show that this excess cannot be explained by means of standard
ionization and recombination processes, since the ratio of the fully
stripped (FeXXVII) to FeXXVI ion fractions becomes (unphysically) negative in the
temperature range of (5 keV, 15 keV), when the observed ratio of
FeXXVI Ly$\gamma$ to FeXXVI Ly$\alpha$ iron lines is applied.

We suggest that a physically acceptable explanation of
the intensity excess in the 8.7 keV line is due to decays of sterile
neutrinos that that are in a halo of dark matter around the Milky Way.
The dark matter density profile we adopt is obtained by
numerical simulations \cite{Klypin 2002}. The dark matter mass from
the Milky Way dark matter halo within the field of view of the
Suzaku observation is small compared with the total mass of the
dark halo $\simeq10^{12} \rm M_{\odot}$, however it suffices
$\simeq 2.5\times10^6 \rm M_{\odot}$ to produce a significant flux
$(1.1\pm0.6)\times10^{-5}$ ph/(cm$^2$ s) via decays of sterile
neutrinos.

{Nearby low surface brightness dwarf galaxies have a central dark
matter density as high as $\langle \rho\rangle \sim 5 \rm
M_\odot/pc^3 (200 \rm GeV/cm^3)$ and are consistent with a common
mass scale of  $\sim 10^7\rm M_\odot$ \cite{Strigari 2008} and even
a universal dark matter profile with a limiting central dark matter
surface density of $\rm log(r_0\rho_0) = 2.15 \pm 0.2, $ in units of
$\rm  log(M_{\odot}/pc^2).$ \cite{Walker 2009}. The corresponding
density is well within the phase space constraints for 17 keV
sterile neutrinos.}

One  explanation of the common
mass scale is that dark matter haloes do not exist  at lower masses.  Warm dark
matter particles have large enough  free streaming lengths
to erase all density fluctuations that might have formed lower mass halos if
the mass of the warm dark matter particle is
approximately 5 keV for a minimum  halo mass of $10^7\rm M_\odot$.
In the present case, the mass of such sterile neutrinos is
$2\times8.7=17.4$ keV. The minimum halo mass scale is approximately
$10^5\rm M_\odot$. Note that the actual dwarf masses could extend to
lower values than evaluated at the 300 pc scale by \cite{Strigari
2008}.

We estimated the mixing angle of
$\sin^2(2\theta)=(4.1\pm2.2)\times10^{-12}$ from the excess in the
intensity and the decay rate is found to be decay rate:
$\Gamma=(9.0\pm4.8)\times 10^{-28} \mathrm{s}^{-1}.$ These values
lie in the allowed region of the mixing angle and decay rate
parameter space for a dark matter sterile neutrino with a mass of
17-18 keV (see \cite{den Herder 2009}). Sterile neutrinos in this
region of mass, decay rate and mixing angle parameter space can
contribute significantly to dark matter. Such neutrinos would be
virtually indistinguishable from cold dark matter except in the
lowest mass dwarf galaxies where cores would be formed. {High
resolution x-ray observations should reveal a point-like line source
of comparable strength to the Suzaku line excess if a cusp is
present around the central black hole.}

\begin{acknowledgements}

We are grateful to Kwang-il Seon and Vladimir Dogiel for valuable
discussions.

\end{acknowledgements}

\end{document}